\documentclass[useAMS,usenatbib]{mn2e}

\include{epsf}

\title[Extended SZ Effect in WMAP Data.]
{Evidence for an Extended SZ Effect in WMAP Data}
\author[A.D. Myers, T. Shanks, P.J. Outram, W.J. Frith and A.W. Wolfendale]
{A. D. Myers\thanks{E-mail:a.d.myers@durham.ac.uk (ADM)},
T. Shanks\thanks{E-mail:tom.shanks@durham.ac.uk (TS)}, P.J. Outram,
W.J. Frith \&  A.W. Wolfendale\\
Dept. of Physics, Univ. of Durham, South Road, Durham DH1 3LE, UK}
\begin{document}

\date{Accepted 2003. Received 2003; in original form 2003 }

\pagerange{\pageref{firstpage}--\pageref{lastpage}} \pubyear{2003}

\maketitle

\label{firstpage}

\begin{abstract}
We have cross-correlated the WMAP data with several surveys of
extragalactic sources and find  evidence for temperature decrements
associated with galaxy clusters and groups detected in the APM Galaxy Survey survey
and  the ACO catalogue. We interpret this as evidence for the thermal
Sunyaev-Zeldovich (SZ) effect from the clusters. Most interestingly, the
signal may extend to $\approx$1 deg ($\approx5h^{-1}\,$Mpc) around
both groups and clusters and we suggest that this may be due to hot
`supercluster' gas. We have further cross-correlated the WMAP data with
clusters identified in the 2MASS galaxy catalogue and also find evidence
for temperature decrements there. From the APM group data we estimate
the mean Compton parameter as $y(z<0.2)=7\pm3.8\times10^{-7}$. We have
further estimated the gas mass associated with the galaxy group and
cluster haloes. Assuming temperatures of 5$\,$keV for ACO clusters and
1$\,$keV for APM groups and clusters, we derive average gas masses of
$M(r<1.75h^{-1}\,$Mpc$)\approx3\times10^{13}h^{-2}M_\odot$ for both, the
assumed gas temperature and SZ central decrement differences approximately 
cancelling. Using the space density of APM groups we then estimate
${\Omega_0}^{gas}\approx0.03h^{-1}$(1$\,$keV$/kT)(\theta_{max}/20')^{0.75}$.
For an SZ extent of $\theta_{max}=20'$, $kT=1\,$keV and $h=0.7$, this
value of ${\Omega_0}^{gas}\approx0.04$ is  consistent with the standard value
of ${\Omega_0}^{baryon}=0.044$ but if the indications we have found for
a more extended SZ effect out to $\theta_{max}\approx60'$ are confirmed, then 
higher values of ${\Omega_0}^{gas}$ will be implied. Finally, the
contribution to the WMAP temperature power spectrum from the extended SZ
effect around the $z<0.2$ APM$+$ACO groups and clusters is 1-2 orders of
magnitude lower than the $l=220$ first acoustic peak. But if a similar
SZ effect arises from more distant clusters then this contribution could
increase by a factor $>10$ and then could seriously affect the WMAP
cosmological fits. \end{abstract}


\section{Introduction}
The Wilkinson Microwave Anisotropy Probe (WMAP) cosmic microwave
background (CMB) anisotropy experiment has published its first year data and
has already provided remarkable results. It has confirmed that the first
peak in the power-spectrum occurs at $l=220\pm10$ and has provided an
excellent further detection of the second peak \citep{wh}. In the first
instance, these new results seem to provide further support for the
standard $\Lambda$CDM cosmology \citep{ws}. But the main new result
from WMAP is the detection of polarisation at large scales which can
only have come from an epoch of reionisation at $10<z<20$ \citep{wk}.
This results in a significant, ($\approx$30 per cent),  reduction of the
acoustic peak heights due to Thomson scattering and shows that the CMB
anisotropies are seriously affected by low redshift galaxy formation
physics. Furthermore, WMAP also finds a low quadrupole in the
temperature power-spectrum \citep{wh} and this is also unexpected if the
Integrated Sachs-Wolfe (ISW) effect caused by relatively recent domination of
the cosmological constant is present. Therefore, although WMAP data
supports the standard model, there are significant problems and already
strong evidence that the temperature spectrum is affected by cosmic
foregrounds such as the re-ionised intergalactic medium at $10<z<20$.

Here we investigate the possibility that other low redshift processes
have filtered the WMAP temperature spectrum. In particular, we shall
search for Sunyaev-Zeldovich (SZ), inverse Compton-scattering of microwave
background photons by hot gas in clusters.  Various authors have made model
dependent predictions for contamination of the CMB data
from the SZ effect and usually concluded that the contaminating effects
were small \citep{refreg1,refreg2,komatsu}. The new WMAP data gives a
first real opportunity to make empirical checks of the level of
contamination by direct cross-correlation of the high resolution
94GHz W band  with galaxy cluster data. The WMAP team
themselves have presented evidence for both SZ and radio source
contamination in the WMAP data. \citet{wb} list 208 point sources
detected at more than $5\sigma$ in the WMAP data and identified them as
radio galaxies and quasars. They also find significant WMAP W band SZ
detections of the brightest X-ray clusters such as Coma and also found a
$2.5\sigma$ detection of the SZ effect in the XBACS sample of 242 X-ray
bright Abell clusters \citep{xbacs}. However, they only looked for SZ
decrements on the scale of the WMAP beam and did not explore any larger
scales.

While this paper was in preparation, several other related  papers have
appeared. \citet{bc} and \citet{wn} have claimed evidence for the ISW
effect in the WMAP data from cross-correlation with the NVSS catalogue.
\citet{diego} have cross-correlated X-ray data and the WMAP data.
Similarly \citet{fg} have cross-correlated APM galaxies and WMAP data;
they find a marginal detection of the ISW effect at 5-10 deg scales
and suggest that the lack of a detection at smaller scales may be due to
cancellation with an SZ effect. \citet{spanish} have obtained upper
limits on diffuse SZ emission from superclusters from failing to find
any cross-correlation with the Abell-Corwin-Olowin (ACO) and other cluster catalogues but they
did find significant correlations from individual clusters. \citet{giommi}
have claimed to detect significant blazar contamination in WMAP and Boomerang data. 
\citet{afshordi} have claimed the detection of SZ, ISW and point sources 
in a power spectrum analysis of WMAP data and the 2MASS galaxy catalogue.

\section{Data}

\subsection{WMAP}

The WMAP data has been published by \citet{wb} in HEALPix format. Here
we shall principally use the WMAP W band data at 94$\,$GHz because of
its relatively high resolution compared to the other bands. The FWHM of
the 94$\,$GHz W beam is $12.'6$ compared to $19.'8$ at V (61$\,$GHz),
$29.'4$ at Q (41$\,$GHz), $37.'2$ at Ka (33$\,$GHz) and  $49.'2$ at  K
(23$\,$GHz). Although, none of the beams is exactly Gaussian (see Fig. 2
of \citet{wp}) we have found that Gaussians of the above FWHM are good
approximate fits to cross-correlation results between faint radio point
sources and WMAP data. We shall also use the Internal Linear Combination
(ILC) map which has 1 deg resolution \citep{wb}. Also the W band has
most sensitivity to the SZ effect via its high resolution, although
perhaps has higher noise than the V band. The HEALPix WMAP data we shall
use has equal area pixels of 49$\,$arcmin$^2$.  Where necessary, we
shall use the Kp0 WMAP mask of \citet{wb} which mainly masks Galactic
contamination but usually its effect is small because we shall generally
be working at Galactic latitudes, $|b|>$40 deg. The maps all use
thermodynamic temperature and the cosmological dipole has already been
subtracted from the data by the WMAP team.

\subsection{Galaxy Cluster Catalogues}

We shall be using three galaxy cluster catalogues. The first is the ACO
catalogue of \citet{aco} which lists clusters with 30 or more members in a
1.5$h^{-1}\,$Mpc radius within $2\,$mag of the 3rd brightest cluster member.
It assigns richness class ($0\leq R\leq 5$) with Coma classed as $R=3$.
The Northern catalogue with $b>40$ deg lists 2489 clusters
with $R\geq0$ the Southern catalogue with $b<40$ deg  lists 1346 such
clusters. The sky density in the North is therefore 0.52$\,$deg$^{-2}$
and in the South it is 0.28$\,$deg$^{-2}$. The sky density of $R\ge2$
clusters is 0.063$\,$deg$^{-2}$ with an average redshift of
$z=0.15$.

We shall also use galaxy group and cluster catalogues derived from the
APM Galaxy Survey of \citet{mad} which covers the whole area with
$\delta<-2.5\,$deg and $b<-40\,$deg. These were identified using the
same `friends-of-friends' algorithm as \citet{adm} and references
therein.  Circles around each APM galaxy with $B<20.5$ are `grown'
until the overdensity, $\beta$, falls to $\beta=8$ and those galaxies
whose circles overlap are called groups. Minimum memberships, $m$, of
$m\geq 7$ and $m\geq 15$ were used which define minimum group
effective `radii' of $1.'2$ and $1.'7$, since the APM galaxy surface
density is $N\approx750\,$deg$^{-2}$ at $B<20.5$. We assume an average
redshift of $z=0.1$ for both APM samples. The sky density of groups
and clusters is 3.5$\,$deg$^{-2}$ at $m\geq7$ and 0.35$\,$deg$^{-2}$
at $m\geq 15$.  Even at $m\geq 15$ there are differences between the
ACO and APM catalogues; an $R=0$ Abell cluster at $z=0.15$ may contain
only a galaxy sky density of 260$\,$deg$^{-2}$ within its $11.'5$
Abell radius compared to a minimum sky density of 5250$\,$deg$^{-2}$
for galaxies within the APM groups.  Therefore the $R\leq 1$ ACO
`clusters' may accommodate much lower density galaxy associations than
even the $m\geq 7$ APM groups which are guaranteed to sample higher
density regions, albeit over smaller areas.

The third cluster catalogue is derived from the final data release of
the 2MASS Extended Source Catalogue (XSC) \citep{jar} to a limit of
$K_s\leq 13.7$. $K$-selected galaxy samples are dominated by early-type
galaxies because of their red colours and early-type galaxies are the
most common galaxy-type found in rich galaxy clusters. Therefore the
2MASS survey provides an excellent tracer of the high density parts of
the Universe out to $z<0.15$ and so provides a further test for the
existence of the SZ effect. Using the above 2-D friends-of-friends
algorithm, we have detected 500 groups and clusters with
$m\ge35$ members at the density contrast $\beta=8$ in the
$|b|\ge10\,$deg area. The 2MASS groups have  average redshift,
$z\approx0.06$.

\section {Cross-correlation Technique}

We search for correlated signals between the 5 ($+$ ILC) WMAP bands and
the above datasets using a simple cross-correlation technique. The WMAP
data are presented in the form of temperature differences with respect
to the global mean and we form the correlation function by calculating
the average $\Delta T$ in annuli around each cluster as a function of
angular distance, $\theta$, between the cluster and pixel centres.

At large $\theta$, where the average background may be expected to be
sampled, the average in particular areas of the sky may not return to
zero, probably because of Galactic contamination. We therefore always
calculate the simple average of all temperature differences in the
sub-area of sky used and plot this on the cross-correlation result as
the effective zero level (solid lines).

The errors are calculated by repeated Monte Carlo random realisations of
the cluster centres across the WMAP area used. The realisations always
have the same number of points as the parent sample and 100 realisations
were done in each case. The effect of clustering of clusters is included
in the realisations; the clustering amplitude is matched to that of the
cluster sample under consideration. In the cross-correlation functions,
there is significant correlation between the results at different
angles, $\theta$, so we have also calculated Monte-Carlo errors in the
integrated bins used to quote overall significances. The Monte Carlo
realisations also quite accurately return results that are close to the
overall average, indicating that our cross-correlation technique is free
of any systematics.  We note that our error analysis will miss variance
from CMB modes of order the cluster catalogue size. Since this size is
generally $\ga50\,$deg, it is not expected that error estimates on
angular scales $\theta\la1\,$deg will be seriously affected, although on
larger scales our quoted errors may increasingly be underestimates.

\section {Cross-Correlation Results}

\begin{figure}
{
\epsfysize=8.4truecm 
\epsfbox[18 144 592 718]{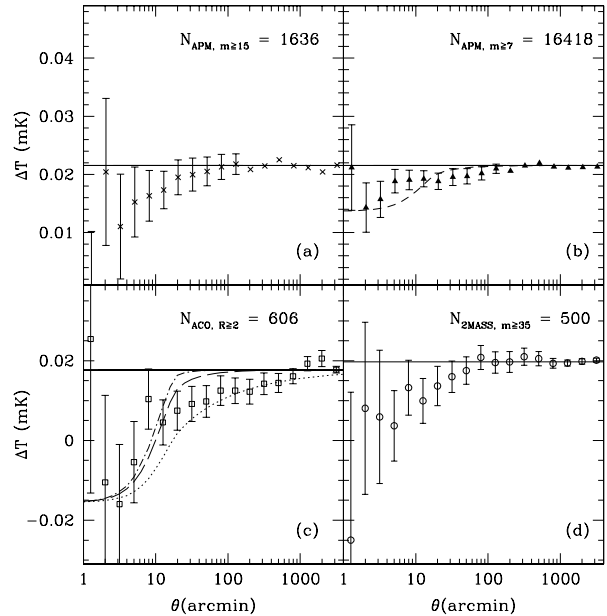}}
\caption{(a) Cross-correlation of WMAP 94$\,$GHz W band data with $m\geq 15$
APM groups and clusters. (b) Cross-correlation of WMAP 94$\,$GHz W band
data with $m\geq 7$ APM groups and clusters. The dashed line is an
isothermal SZ model with $\Delta T(0)=0.015\,$mK, $\theta_c=2.'3$,
$\beta=0.75$ convolved with a Gaussian beam profile with $\sigma=5.'25$.
(c) Cross-correlation of WMAP 94$\,$GHz W band data with ACO $R\geq 2$
clusters for combined ACO $|b|>40\,$deg N$+$S samples. The dashed,
dotted and dot-dash lines are isothermal SZ models with $\theta_c=1.'5$
and $\Delta T(0)=0.083\,$mK, $\beta=0.75$; $\Delta T(0)=0.050\,$mK,
$\beta=0.5$; and $\Delta T(0)=0.12\,$mK, $\beta=1.0$, convolved as in
(b). (d) Cross-correlation of WMAP  94$\,$GHz W data with 500 galaxy
groups and clusters selected by friends-of-friends
algorithm ($\beta=8$, $m\ge35$) from the $K_s<13.7$ 2MASS XSC. In all
cases the solid lines represent the average $\Delta T$ obtained over the
area used. Monte-Carlo errors including the effects of clustering of clusters
are shown.}
\end{figure}

In Fig. 1a,b we first present the results from cross-correlating our
APM groups and cluster centres with the WMAP W band data. We consider
the two APM cluster samples with $m\geq15$ and $m\geq7$ members. Both
samples show indications of anti-correlation with respect to the
94$\,$GHz data. Relative to the overall mean, the integrated
significances for the $m\geq15$ sample range from $1.7\sigma$ at
$\theta<10'$ to $1.0\sigma$ at $\theta<30'$. The significances for the $m\geq7$
sample range from $2.0\sigma$ at $\theta<10'$ to $1.8\sigma$
at $\theta<30'$.  In the case of the $m\geq15$ clusters, the size of
the anti-correlation is $\Delta T\approx -0.01\,$mK on WMAP pixel
scales.  It is also interesting that the size of the signal only goes
down slightly to $\Delta T\approx -0.008\,$mK at small scales for the
$m\geq 7$ sample and here the anti-correlation appears to extend
beyond the beam-size, only reducing to $1.5\sigma$ at
$\theta\approx 60'$. The anti-correlation signal seen at small
$\theta$ ($\leq 10'$) is consistent with what is expected at this
frequency for the temperature decrement caused by the SZ effect
\citep{refreg1}. The surprise is that it can be seen out to
$\approx$1$\,$deg scales and that it persists in groups with a sky
density of 3.5$\,$deg$^{-2}$, a factor $\approx$30$\times$ higher than
that of the Abell clusters.

We now turn to the ACO catalogue to see if there is any confirmation of
this tentative detection of extended anti-correlation around the APM
groups and clusters. When the full ACO catalogue at $|b|>40\,$deg
including clusters of all richnesses, was used only an insignificant
anti-correlation was found. The catalogue was then cut back to $R\geq2$
and the result changed dramatically and consistently in both
Hemispheres. Based on the remaining 229 clusters with $b>40\,$deg and
the remaining 377 clusters with $b<-40\,$deg, significant
anti-correlation is again seen out to  scales of
$\theta\approx1\,$deg or $\approx$7.5$h^{-1}\,$Mpc (see Fig. 1c). The
Monte-Carlo errors indicate that the effect is significant at the
$\approx2.1\sigma$ level for $\theta<10'$ and at the
$\approx2.2\sigma$ level for $\theta<1\,$deg in the N$+$S sample.
Our interpretation is that the $R=0,1$ clusters may have too low
densities to produce a strong SZ effect. We conclude that there is a
significant anti-correlation out to $\approx1\,$deg scales between the
ACO $R\geq 2$ cluster catalogue and the WMAP W band data, confirming the
tentative detection seen in the APM group and cluster catalogues, and
that this is likely to be caused by the SZ effect.

\begin{figure}
{
\epsfysize=8.4truecm 
\epsfbox[18 144 592 718]{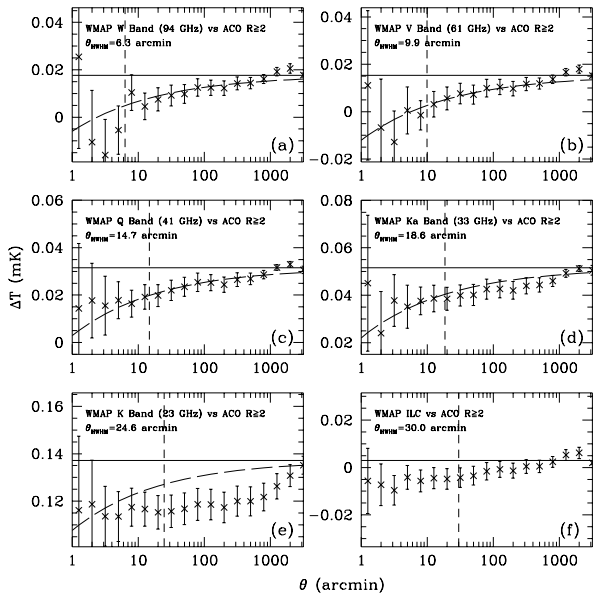}}
\caption{Cross-correlation of WMAP data in the W, V, Q, Ka, K and ILC
bands with 606 $|b|>40\,$deg ACO clusters. The SZ decrement can be seen
in each band.  In all cases the solid line is the simple average $\Delta
T$ over the area surveyed, and the vertical short-dashed line indicates
the beam half-maximum in each band.  The long-dashed line shows a $\theta^{-1/3}$ fit
to the W band, scaled for the SZ frequency dependence to the
other bands,  which can be used as a reference line for the frequency
dependence. }
\end{figure}

Figs. 2 (a-f) shows that the ACO $R\geq 2$ clusters also produce
anti-correlations in all the other frequency  bands, including the ILC.
In terms of the predicted SZ frequency dependence according to equations
(11) and (13) of \citet{refreg1} this is expected because relative to
the W band the V, Q, Ka and K bands the SZ decrement should increase by
the factors 1.16, 1.21, 1.25 and 1.25. The poorer resolution of the low
frequency WMAP bands  makes the detection of the decrements more
unexpected but it is explained by the apparent persistence of the signal out to
scales well in extent of the W band beam. The frequency dependence of
the SZ signal appears statistically consistent with the SZ prediction
(dashed lines) in all except the K band where the combination of poor
resolution and residual Galactic contamination may be the cause of the
poor fit. However,  the errors are too large to discriminate between the
SZ and CMB spectral indices. It is this spectral similarity which makes
it difficult to reject SZ contamination in the CMB maps and which
explains our SZ detection in the `clean' ILC map. The APM clusters and
groups also show anti-correlations in the V and Q bands but less so in
the other bands, probably because of the lower resolution. 

To check if the cross-correlation results contain artefacts
characteristic of residual Galactic foreground contamination, we have
cross-correlated the ACO clusters with the WMAP foreground maps of
\citet{wb} and find no indication of any strong systematics. We also
note that we have obtained results for the SZ decrements from the ACO,
APM and 2MASS clusters from the `foreground-cleaned' map of
\citet{toch} that are entirely consistent with those found from the
ILC map; the detection of the same anti-correlation in both of these
`clean' maps further argues against the possibility that the
anti-correlation is due to foreground systematics.

\begin{figure}
{
\epsfysize=8.4truecm 
\epsfbox[18 144 592 718]{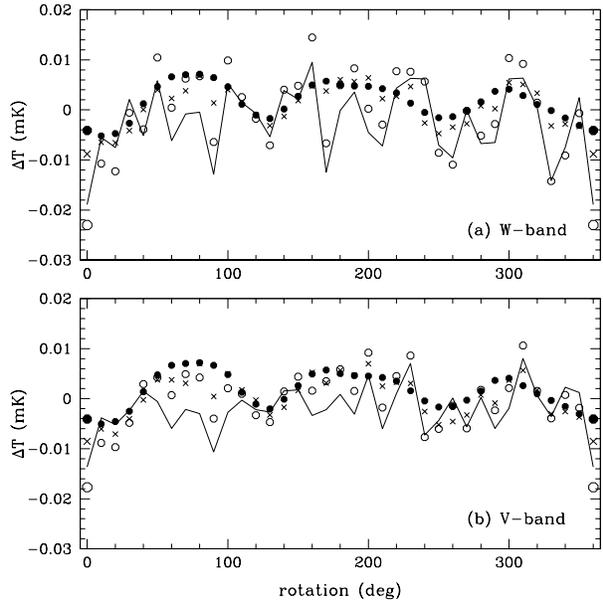}}
\caption{(a) Cross-correlation of WMAP W band  with the 606 ACO clusters
after rotation about the Galactic Poles by adding $\Delta l$ to the
cluster's Galactic longitude.  The open circles represent the  average
$\Delta T$ with $\theta<6.'3$ (the beam half-maximum in W), the crosses
represent the average $\Delta T$ with $\theta<60'$ and the filled
circles represent the average $\Delta T$ with $\theta<500'$. In each
case, the average $\Delta
T$ over the area surveyed has been subtracted. The
$\theta<6.'3$ and $\theta<60'$ results at zero rotation lie
significantly below the others but the equivalent $\theta<500'$ result 
does not. The solid line represents the difference between the $\theta<6.'3$ 
and the $\theta<500'$ results which shows the effect on the beam-size result 
if the anti-correlation observed at $\theta<500'$ is assumed due 
entirely to systematics. (b) As (a) for the V band. Here the open circles 
and solid line correspond to $\theta<9.'9$, the beam half-maximum in V.}
\end{figure}

As a further powerful check against systematics caused by foreground
contamination, we have also applied the cross-correlation analysis after
rotating the ACO $R\geq 2$ clusters around the Galactic Poles with
respect to the WMAP W and V band data. Fig. 3 shows the W and V band
cross-correlation signals (relative to the mean $\Delta T$) integrated
to the beam half-maximum ($\theta<6.'3$ for W, $\theta<9.'9$ for V),
$\theta<60'$ and $\theta<500'$ measured at intervals of $\Delta
l=10$ deg of Galactic longitude. The solid line shows the difference
between the beam half-maximum  and the $\theta<500'$ angular scales;
this is to test for the persistence of the beam-size anti-correlation
under the assumption that the $\theta<500'$ result is entirely due to
systematic effects. The anti-correlation at the beam half-maximum is
clearly significant whether measured relative to the overall mean
($3.1\sigma$ for W, $3.6\sigma$ for V) or the $\theta<500'$ results
($2.8\sigma$ for W, $3.1\sigma$ for V). The anti-correlation at
$\theta<60'$ also remains significant whether measured relative to the
overall mean ($2.3\sigma$ for W, $2.2\sigma$ for V) or relative to the
$\theta<500'$ results ($2.0\sigma$ for W, $1.7\sigma$ for V). Although
the rotated results appear more correlated at $\theta<60'$ than at
$\theta<6.'3$, implying these latter significances be treated with
caution, it remains the case that for the $\theta<60'$ the zero rotation
result shows the lowest $\Delta T$. This is not the case for the
$\theta<500'$ points where several of the rotated points show both lower
and higher excursions around zero than the zero rotation result. Clearly
there may be systematics which are beginning to dominate any real SZ
signal at these very large scales. We conclude that the rotation
experiments suggest that the observed temperature decrements around ACO
clusters are significant from the beam-size out to $\theta<60'$ and
these results are robust to the assumption that the cross-correlation is
dominated by possible systematics on $\theta<500'$ scales.

Fig. 1(d) shows the cross-correlation of the 500 2MASS groups and
clusters limited at $K_s<13.7$ with the  WMAP  W band. Again  the 2MASS
data shows similar trends to the APM and ACO clusters with
anti-correlation seen on scales of the beam and extending at
marginal significance out to scales of $\approx1$ deg.  Again the
frequency dependence is consistent with an SZ signal (not shown). At the
average 2MASS group redshift, $z\approx0.06$, $\theta<1$ deg
corresponds to $r<3h^{-1}$Mpc and thus here there is further evidence
for possible extension of the SZ decrements to scales of order
$>1h^{-1}$Mpc.

\section {Discussion}

We have found evidence for WMAP temperature decrements extending out to scales of
$\theta\approx1$ deg from galaxy groups and clusters, with hints that
the decrements may extend to even larger scales. Previous
cross-correlation of the ACO catalogue with the Rosat All-Sky Survey has
shown  diffuse X-ray emission from $\approx1\,$keV gas extending out to
$\approx2$ deg from $D=5$, $R\geq 1$ Abell Clusters, comparable to the
scale of the anti-correlation seen in Fig. 1c \citep{soltan}. Thus,
based on this indication that the anti-correlation may be caused by the
SZ effect originating from diffuse `supercluster' gas, we make a first
order calculation of the overall Compton parameter, $y$. From
\citet{refreg1} we use the relation $\Delta T_{SZ}/T_0=yj(x)$
where $T_0$ is the CMB temperature, $x=h\nu/kT_0$ and $j(x)$
is a spectral function which takes the value $j(x)=-1.56$ at
94$\,$GHz.  The simplest route is to go via the $m\geq 7$ APM groups
and clusters because these have the biggest space density and so are
the most representative of average sightlines. These have a sky
density of 3.5$\,$deg$^{-2}$ with an average SZ decrement that extends
to $\theta>0.5\,$deg and so have a sky covering factor of
approximately unity. For $0.1<\theta<0.5\,$deg in Fig. 1b, $\Delta T$
is reasonably flat so that effects due to the WMAP beam may be small
and in this range $\Delta T_{SZ}= -3.0\pm1.6\,\mu$K. Using the above
relation with $T_0=2.726\,$K, this converts into a value for
$y(z<0.2)=7\pm3.8\times10^{-7}$. Now \citet{refreg1} refer to
\citet{scar} and \citet{persi} as suggesting that 40 per cent of $y$
originates at $z<0.2$ in CDM models. On this assumption, we find
$y(z<\infty)=1.8\pm1.0\times10^{-6}$. This compares to the 3$\sigma$
upper limit on the total integrated $y$ parameter from the COBE-FIRAS
measurement of the spectral distortion of the CMB of
$y(z<\infty)=2.2\times10^{-5}$ \citep{fix1}. The 3$\sigma$ upper limit
from cross-correlating COBE DMR and FIRAS is
$y(z<\infty)=4.5\times10^{-6}$ \citep{fix2}. \citet{banday} found a
3$\sigma$ upper limit of $\delta y(z<0.2) <1.5\times10^{-6}$ by
cross-correlating COBE DMR with the ACO cluster catalogue. Thus our
result is not inconsistent with these previous observational upper
limits. We also note that our estimate of $y$ is 2-3$\times$ higher
than that predicted in the SCDM model of \citet{scar} and that it is
similar to the value predicted in the $\Lambda$CDM model of
\citet{persi}.

We next fit isothermal models to the SZ decrements for the ACO R$\ge$2
clusters. The models are taken  from Equns. 15, 16 of \citet{refreg1}
and convolved with Gaussians to represent the beams. We 
assume the value of $\beta=0.75$ quoted for Coma and a value of
$\theta_c=1.'5$ which is the Coma value scaled approximately to the
$z=0.15$ average redshift of the ACO sample. We find that 0.083$\,$mK is
the best fit for $\Delta T(0)$ compared to the 0.5$\,$mK quoted for
Coma. The results are shown in Figs. 1(b),(c). It can be seen that the
data appears to show a more extended decrement than the model. The same
model with $\beta=0.5$ gives an improved fit at larger scales for
$\Delta T(0)$ of 0.05$\,$mK. 

The cluster correlation function \citep{bs} suggests the number of
excess clusters at $20'<\theta<100'$ from an average cluster is
$\approx1.3$. Based on the average decrement seen at $\theta<20'$, we
estimate this cluster excess will contribute $\Delta T=0.5\,\mu$K, at
$20'<\theta<100'$, compared to the observed $\Delta T=6\,\mu$K.  Thus
it does not seem possible for clustering of `beam-sized' clusters to
explain the extent of the decrement; an extended ($\theta\ga1$ deg)
gas halo around individual clusters appears to be needed.

We next calculate the gas mass associated with an ACO $R\ge2$ cluster.
Using equation (16) of \citet{refreg1} and assuming $kT=5\,$keV,
$r_c=0.2h^{-1}\,$Mpc, the W band fit with $\Delta T(0)=0.083\,$mK gives
a value for the central electron density, $n_0=1.8h\times
10^{-3}\,$cm$^{-3}$. Integrating equn. (14) of \citet{refreg1} with
$\beta=0.75$ to $r<1.75h^{-1}\,$Mpc ($\approx 13'$) then gives a gas
mass of $M\approx3\times10^{13}h^{-2}M_\odot$, which is not unreasonable
compared to the X-ray gas mass of
$M\approx1\times10^{14}h^{-2.5}M_\odot$ detected within a similar radius
in the Coma cluster \citep{lea}.

A model with $\beta=0.75$, $\theta_c=2.'3$ and $\Delta T(0)=0.015\,$mK
is shown to give a reasonable fit at small scales to the APM $m\ge7$
sample in Fig. 1c. $\theta_c=2.'3$ is the Coma value of
$r_c=0.2h^{-1}\,$Mpc scaled to $z=0.1$. Again there is evidence that
the profile is more extended than the model.

We now proceed to calculate the average gas mass associated with an
APM $m\ge7$ group/cluster. Using equation (16) of \citet{refreg1} and
assuming $r_c=0.2h^{-1}\,$Mpc, the W band fit with $\Delta
T(0)=0.015\,$mK gives a value for the central electron density,
$n_0=1.6h\times 10^{-3}\,$cm$^{-3}(kT/1\,$keV). Integrating equn. (14)
of \citet{refreg1} with $\beta=0.75$ to $r<1.75h^{-1}\,$Mpc
($\approx20'$) then gives a gas mass of
$M\approx3\times10^{13}h^{-2}(kT/1\,$keV)$M_\odot$.  Since the APM groups
and clusters are more numerous than the ACO clusters, we choose to use
them to make an estimate of ${\Omega_0}^{gas}$. Assuming the space
density of APM groups quoted by \citet{cs,adm} of
$3\times10^{-4}h^3\,$Mpc$^{-3}$ we find a value of
${\Omega_0}^{gas}\approx0.03h^{-1}(1\,$keV$/kT)(\theta_{max}/20')^{0.75}$.
Assuming $h=0.7$, $kT=1\,$keV and $\theta_{max}=20'$ gives
${\Omega_0}^{gas}\approx0.04$, close to the WMAP
${\Omega_0}^{baryon}=0.044\pm0.004$ result \citep{ws}. There may be
evidence in Fig. 1(b) for the SZ decrements to extend to
$\theta_{max}\approx 60'$.  In this case, the masses associated with
APM groups rise by a further factor of $\approx2$ implying
${\Omega_0}^{gas}\approx0.1$, now a factor of $\approx2\times$ higher than
the standard value for ${\Omega_0}^{baryon}$.

Is there any possibility that SZ decrements in the WMAP 94$\,$GHz data
could contaminate the acoustic peaks measurement? The question is
interesting because the SZ correlations appear to extend out to
$\theta\approx1$ deg, the location of the first acoustic peak in the power
spectrum. We have therefore run some simple models where circular
areas representing clusters with SZ decrements are distributed at
random over simulated CMB fields, laid down from Gaussian random
fields drawn from CMBFAST power spectra \citep{zs}. If the SZ clusters
have the same 3.5$\,$deg$^{-2}$ sky density as the APM $m\geq 7$
groups and we assume that they extend only to $\theta<0.5\,$deg with a
decrement $\Delta T_{SZ}=-3\,\mu$K then the SZ power spectrum of
clusters is two orders of magnitude below the first peak at wavenumber
$l=220$. However, if we assume that these groups and clusters extend
unevolved in their gas content past the $z<0.2$ APM limit out to
$z<0.5$, then the group $+$ cluster sky density will rise to
$\approx50\,$deg$^{-2}$. If we further assume values for the SZ extent
and decrement which are at the high end of the ranges so far
suggested, i.e. $\theta<1\,$deg and $\Delta T_{SZ}=5\,\mu$K, then an
SZ peak appears in the power spectrum which is of order $\approx$30
per cent of the amplitude of the first acoustic peak at $l=220$. The
consequences for the lower amplitude acoustic peaks at higher
multipoles could be even more serious; we note that the temperature
power spectrum from the CBI experiment already shows an excess at
$l>2000$ which may be due to a strong SZ effect from individual
clusters \citep{cbi,js}. The above models might run up against the
$y(z<\infty)=4.5\times10^{-6}$ upper limit of \citet{fix2}. {\it
However, it is clear that the question of how much SZ effects
contaminate the primordial power spectrum is re-opened by the spatial
extent of the SZ signal found in our results.}  Higher resolution CMB
data and deeper group and cluster catalogues are
needed to constrain further the SZ contribution from $z>0.2$ clusters.

\section {Conclusions}

We have found evidence for anti-correlation between WMAP data and the
positions on the sky of galaxy clusters derived from the ACO, APM and
2MASS surveys. We interpret the signal as caused by the SZ effect,
inverse Compton scattering of the CMB photons by hot gas in groups and
clusters of galaxies. The signal may extend to $\approx1\,$deg scales
around ACO clusters, implying they have extended gaseous haloes which
may also constitute a diffuse gas component in superclusters. We
estimate the mean Compton $y$ parameter associated with $z<0.2$ APM
groups and clusters as $y(z<0.2)=7\pm3.8\times10^{-7}$. This is not
inconsistent with previous upper limits and with expectations from CDM
models. We have also estimated the average $R\ge2$ ACO cluster gas
mass assuming $kT=5\,$keV and found this to be in reasonable agreement
with X-ray observations of individual ACO clusters within the central
radius, $r<1.75h^{-1}\,$Mpc. The average mass of APM groups$+$clusters
is found to be $M(r<1.75h^{-1}\,$Mpc$)\approx3\times10^{13}h^{-2}(1\,$keV$/kT)M_\odot$
which gives ${\Omega_0}^{gas}\approx0.03h^{-1}(1\,$keV$/kT)$. For
$kT=1\,$keV and $h=0.7$ this value for ${\Omega_0}^{gas}\approx0.04$ is
close to the value of ${\Omega_0}^{baryon}=0.044$ in the standard
model \citep {ws}.  But because the X-ray temperatures are likely to
be lower than $1\,$keV and because the SZ decrements show evidence for
gas haloes which extend beyond $r=1.75h^{-1}\,$Mpc then this value for
${\Omega_0}^{gas}$ could well prove to be a lower limit on the true
value.

We have briefly investigated how SZ contamination might affect the
location and shape of the acoustic peaks in the WMAP temperature and
find that although there is little effect from the temperature
decrements found so far, if they persist out to $z\approx0.5$ with
the amplitude and extent seen at $z<0.2$, then even the first acoustic
peak at the  $1\,$deg scale could be significantly affected. Further
cross-correlation analysis of deeper catalogues of groups and clusters
will be needed to judge the seriousness of this potential SZ
`contamination'.

\section*{Acknowledgments} We thank A.J. Banday for useful discussions.
A.D. Myers and W.J. Frith acknowledge receipt of PPARC PhD studentships.
P.J. Outram acknowledges receipt of a PPARC fellowship. We thank an
anonymous referee for very useful comments on earlier versions of this
paper.

\label{lastpage}

\end{document}